\documentclass[runningheads]{llncs}

\usepackage[utf8]{inputenc}
\usepackage{times}
\usepackage{soul}
\usepackage{url}
\usepackage[small]{caption}
\usepackage{amssymb}
\usepackage{amsmath}
\usepackage{mathtools}
\usepackage{booktabs}
\usepackage[pdf]{graphviz}
\usepackage[linesnumbered,algoruled,boxed,lined,noend]{algorithm2e}
\usepackage{tikz}
\usepackage{todonotes} 
\usepackage[hidelinks]{hyperref} 

\usetikzlibrary{automata, arrows, calc, patterns, positioning}
\tikzset{auto, >=stealth}
\tikzset{every edge/.append style={shorten >=1pt}}

\newcommand{\Lstar}{L\textsuperscript{$\ast$}}
\newcommand{\PLSynth}{\Lstar-PSynth}

\newif\ifdraft\drafttrue
\ifdraft
\newcommand{\anthony}[1]{\color{red} {AL: #1 :LA} \color{black}}
\newcommand{\oliver}[1]{\color{brown} {OM: #1 :LZ} \color{black}}
\newcommand{\najib}[1]{\color{blue} {NM: #1 :LT} \color{black}}
\newcommand{\chihduo}[1]{\color{cyan} {CH: #1 :HM} \color{black}}
\newcommand{\daniel}[1]{\color{magenta} {DN: #1 :PR} \color{black}}
\else
\newcommand{\anthony}[1]{}
\newcommand{\oliver}[1]{}
\newcommand{\najib}[1]{}
\newcommand{\daniel}[1]{}
\newcommand{\chihduo}[1]{}
\fi

\newcommand{\OMIT}[1]{}

\author{Oliver Markgraf\inst{1} \and
	Chih-Duo Hong\inst{3} \and
	Anthony W. Lin\inst{1,2}\and
	Muhammad Najib \inst{1} \and 
	Daniel Neider \inst{2}
}
\institute{Technical University of Kaiserslautern, Germany \and
	Max Planck Institute for Software Systems, Kaiserslautern, Germany \and 
		University of Oxford, England\\
}
\begin{document}
\title{Parameterized Synthesis with Safety Properties}
%
%
%
%
\maketitle              
\begin{abstract}
Parameterized synthesis offers a solution to the problem of constructing correct and verified controllers for parameterized systems.
Such systems occur naturally in practice (e.g., in the form of distributed protocols where the amount of processes is often unknown at design time and the protocol must work regardless of the number of processes).
In this paper, we present a novel learning-based approach to the synthesis of reactive controllers for parameterized systems from safety specifications.
We use the framework of regular model checking to model the synthesis problem 
    as an infinite-duration two-player game and show how one can utilize Angluin's well-known \Lstar\ algorithm to learn correct-by-design controllers.
This approach results in a synthesis procedure that is conceptually simpler than
    existing synthesis methods with a completeness guarantee,
    whenever a winning strategy can be expressed by a regular set.
We have implemented our algorithm in a tool called \PLSynth\ and have demonstrated its performance on a range of benchmarks, including robotic motion planning and distributed protocols.
    Despite the simplicity of \PLSynth\, it competes well against (and in many 
    cases even outperforms) the state-of-the-art tools for synthesizing 
    parameterized systems.
\end{abstract}
%
%
%


\section{Introduction}
\label{sec:intro}
Parameterized systems are systems with a parameterized 
number of components. Such systems are ubiquitous in distributed and/or
reactive systems, (e.g.,
where the number of clients, the size of the environment, etc.\ can take
arbitrary finite values and the correctness property must hold regardless of
the assigned value). For example, in order to verify safety/liveness of a Dining 
Philosopher Protocol with $n$ philosophers, we need to prove the property
for \emph{each} value of $n \geq 3$. This is known as the \emph{parameterized
verification problem},
which is undecidable even for safety properties \cite{APT86}.

Verification of parameterized systems has been the subject of many papers
spanning across four decades (e.g., see
\cite{sasha-book,Parosh12,ZP04,vojnar-habilitation} for
surveys). Many different techniques for verifying
parameterized systems have been proposed including cutoff techniques~\cite{sasha-book,AHH16},
acceleration~\cite{Parosh12,RMC-tree}, learning~\cite{LR16,CHLR17,NT16,Lever,Lever-omega}, and abstractions~\cite{abstract-RMC},
to name a few. The problem of verifying \emph{safety} property (i.e., bad things will 
never
happen) has occupied a lot of these research results, owing to its widely
recognized importance. 

In this paper, we are interested in automatically synthesizing correct 
parameterized systems with a safety guarantee. In this setting, parameterized
systems are only partially specified, and the task of a synthesis algorithm
is to ``fill in'' the missing specification in such a way that the desired
property is satisfied. Synthesis algorithms aim to produce a 
correct-by-construction implementation of some formal properties in a fully 
automatic fashion, thereby saving the need for performing a further verification
step. Program synthesis has been an active research area with many applications
(e.g., 
to patch faulty parts of a system~\cite{SB07,JGB05,GSB07,FSBD08} or to fill the low-level details of a partial 
implementation~\cite{SL09,LTBSS06,LATBSS07}).
However, there has not been much work 
on
synthesis for parameterized systems with safety guarantee. 


A common approach to the synthesis with a safety guarantee is by utilizing 
games, more specifically a 
type of games called \textit{safety games}.
Safety games are two-player games with \textit{safety objectives} (i.e., the objective is to
always stay inside a ``safe'' region). Safety games have been widely applied in the
context of verification and synthesis of reactive systems. One example of their 
usage is for synthesis of safe controllers, such as a vacuum cleaner robot that tries to avoid bumping into humans while cleaning the room or a controller for a safety-critical system that
maintains the temperature of a power plant within a certain safe level.
Safety games have been extensively studied in many settings in the literature, both with finite-state arenas and infinite-state arenas, and including timed 
systems, hybrid systems, counter systems, and arenas generated by finite-state
transducers. Some examples, among many others, can be found in
\cite{gradel2002automata,NT16,NM19,CMBM18,tomlin,doyen11,ESK14,CDL09}.
A parameterized system can naturally be construed as an infinite-state system.
Each parameter instantiation gives us a finite system,
but there are infinitely many such instantiations. The corresponding 
infinite-state system is a disjoint union of all 
finite systems obtained from all possible parameter instantiations.
This is an undecidable problem; in fact, verifying safety properties (i.e. 
one-player games) is already undecidable for parameterized systems \cite{APT86}.
There are a handful of generic methods and tools that have been designed in the 
past six years to handle safety games over general infinite-state systems 
\cite{BCPR14,NT16,Katis18,NM19}.
Examples include CONSYNTH~\cite{BCPR14}, DT-Synth~\cite{NM19}, JSyn-VG~\cite{Katis18}, SAT-Synth~\cite{NT16}, and RPNI-Synth~\cite{NT16}, which have varying degrees of automation and expressivity.
For instance, the former three synthesis tools (i.e., CONSYNTH, DT-Synth, and JSyn-VG) support safety games over arenas with infinitely many vertices that are modeled using integer or real linear arithmetic.
By contrast, the latter two tools (i.e., SAT-Synth and RPNI-Synth) work in a setting similar to \emph{regular model checking} \cite{Parosh12,rmc-pnueli}, which encodes parameterized systems by means of regular languages and finite-state transducers.
Since regular model checking is a popular and highly expressive framework for modelling and verifying parameterized systems, we follow the approach by SAT-Syth and RPNI-Synth throughout this paper.

Many of these aforementioned algorithms rely heavily on user guidance or are highly intricate.
CONSYNTH, for instance, requires the user to provide templates that carry high-level information about possible solutions in order to prune the search space.
SAT-Synth, on the other hand, repeatedly solves an NP-complete problem (learning of minimal finite-state machines from examples) and, hence, is computationally expensive.
In this paper, we thus provide a different and \emph{substantially simpler} solution to the synthesis problem, which does not require user guidance and is computationally efficient.

\paragraph*{Contribution.}
The main contribution of this paper is to show how a simple \emph{exact learning} algorithm for
automata (e.g. Angluin's $L^*$ algorithm \cite{angluin1987learning}) can be employed 
effectively for solving regular safety games in regular model 
	checking \cite{Parosh12}, while remaining competitive with existing tools
for parameterized synthesis with safety properties.
Furthermore, we
show the efficacy of our procedure in various problem domains including path 
planning in a grid with adversaries, two-player zero-sum games (e.g. Nim), and 
distributed protocols. We elaborate below why
this is a challenging problem.

We first quickly recall the framework of exact learning of regular languages
\cite{angluin1987learning,learning-book}. A learner's goal is to learn 
an unknown
regular language $L$ (represented by minimal DFA --- deterministic finite 
automaton)
with the guide of a teacher, who can answer a \emph{membership} query and an
\emph{equivalence} query. A membership query checks whether a given word $w \in
\Sigma^*$ is in $L$. On the other hand, an equivalence query asks whether
the language $L' := L(A)$ of a given DFA $A$ coincides with $L$; if not, the teacher has to
return a counterexample $w \in (L\setminus L') \cup (L'\setminus L)$ to the
learner. In her seminal paper \cite{angluin1987learning}, she provided the so-called L*
algorithm, which learns a DFA in polynomial-time\footnote{The running time by
	definition accounts for the amount of time taken by the learner plus the maximum
	size of the counterexamples provided by the teacher. We assume the teacher is an
	oracle that can return an answer in constant time.}. Different exact learning
algorithms for automata are by now available that in practice may outperform
Angluin's original algorithm, e.g., see \cite{learning-book}. 

Angluin's exact learning of regular languages is conceptually simple, but
when a problem can be successfully modelled in this framework (e.g. 
see \cite{chen-clarke,CHLR17} for such examples in verification), one
can tap into a wealth of efficient learning algorithms. 
When employing this for infinite-state
verification, the language $L$ to be learned typically represents a kind of
correctness proof (e.g. invariants). 
This is problematic because this is 
\emph{not unique}, which is necessary for a successful modelling in the exact
learning framework. 
The proposed strategy in this paper is to design the
so-called \emph{strict but generous teacher}, which essentially drives the
learner to learn the safe region reachable from the set of initial states (which
is \emph{unique}) but accepts a different correct proof from the learner. 
For this idea
to work, a membership query (asking whether a given configuration is reachable
and in a safe region) should not be an undecidable problem. 
To this end, we propose to consider length-preserving transducers, which is 
known to be sufficiently general \cite{Parosh12}. With this restriction,
we obtain a framework where membership queries become decidable, and
can in fact be checked using fast finite-state model checkers.

We have implemented our approach in a tool called \textit{\PLSynth}. We also provide some case studies as benchmarks in order to evaluate our implementation. Some of the case studies are taken from \cite{NT16}, while the rest are known games, or inspired by some real world applications. Furthermore, we compare the performance of our tool (using the provided benchmarks) against three existing sate-of-the-art tools: \textit{SAT-Synth}, \textit{RPNI-Synth}~\cite{NT16} and \textit{DT-Synth}~\cite{NM19}. Despite its simplicity, the tool competes well in practice against the other three tools, and even in many cases, outperforms them.

\paragraph*{Organization.} We start with a couple of motivating examples in the next section. Section~\ref{sec:preliminaries} contains preliminaries. We describe the algorithm of our proposed approach in Section~\ref{sec:algo}. In Section~\ref{sec:experiments}, we provide some case studies and report the experiments to measure the performance of our implementation against two existing tools. We conclude in Section~\ref{sec:conclusion}.


\section{Motivating Examples}
\label{sec:examples}
\paragraph*{Robotic motion planning example.}
Consider two robots inhabiting a bounded two-dimensional grid world, one controlled by a controller/system that we wish to synthesize, and the other controlled by the environment (which we do not control.) 
We call this game ``follow game'', which, later in Section~\ref{sec:experiments}, is also used as one of the benchmarks. In this game, both robots move in alternating turns, and by one grid on each turn.
The goal of the game is to find (and synthesize) a strategy such that the robot controlled by the system stays within a certain distance to the environment's robot. We can consider this game as an abstraction of some system in which some drones need to be in close proximity to some moving targets. Such a strategy thus can be synthesized as a controller for the drones. 

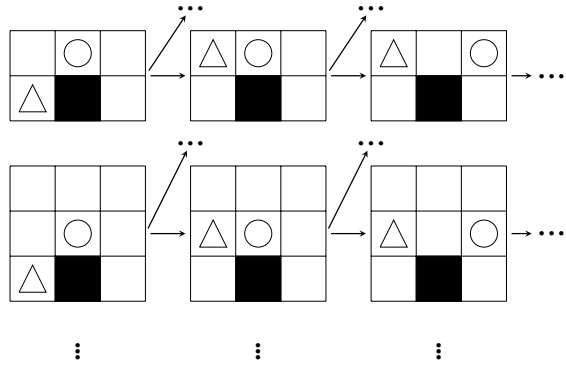
\begin{figure}[t]
\centering
\scalebox{0.6}{
\begin{tikzpicture}
\draw (0,0) grid (3,2);
\draw (0.2,0.2) -- (0.8,0.2) -- (0.5,0.8)-- (0.2,0.2);
\draw (1.5,1.5) circle (0.3);
\draw[fill = black] (1,0) rectangle (2,1);
\node (a) at (3,1){};
\node (b) at (4,1){};
\node (c) at (4,2.5){\Huge ...};
\draw [thick, ->] (a) to (b);
\draw [thick, ->]  (a) to (c);

\draw (4,0) grid (7,2);
\draw (4.2,1.2) -- (4.8,1.2) -- (4.5,1.8)-- (4.2,1.2);
\draw (5.5,1.5) circle (0.3);
\draw[fill = black] (5,0) rectangle (6,1);
\node (c) at (7,1){};
\node (d) at (8,1){};
\node (e) at (8,2.5){\Huge ...};
\draw [thick, ->] (c) to (d);
\draw [thick, ->] (c) to (e);

\draw (8,0) grid (11,2);
\draw (8.2,1.2) -- (8.8,1.2) -- (8.5,1.8)-- (8.2,1.2);
\draw (10.5,1.5) circle (0.3);
\draw[fill = black] (9,0) rectangle (10,1);
\node (e) at (11,1){};
\node (f) at (12,1){\Huge ...};
\draw [thick, ->] (e) to (f);

\draw (0,-4) grid (3,-1);
\draw (0.2,-3.8) -- (0.8,-3.8) -- (0.5,-3.2)-- (0.2,-3.8);
\draw (1.5,-2.5) circle (0.3);
\draw[fill = black] (1,-4) rectangle (2,-3);
\node (a) at (3,-2.5){};
\node (b) at (4,-2.5){};
\node (c) at (4,-0.5){\Huge ...};
\draw [thick, ->] (a) to (b);
\draw [thick, ->] (a) to (c);

\draw (4,-4) grid (7,-1);
\draw (4.2,-2.8) -- (4.8,-2.8) -- (4.5,-2.2)-- (4.2,-2.8);
\draw (5.5,-2.5) circle (0.3);
\draw[fill = black] (5,-4) rectangle (6,-3);
\node (c) at (7,-2.5){};
\node (d) at (8,-2.5){};
\node (e) at (8,-0.5){\Huge ...};
\draw [thick, ->] (c) to (d);
\draw [thick, ->] (c) to (e);

\draw (8,-4) grid (11,-1);
\draw (8.2,-2.8) -- (8.8,-2.8) -- (8.5,-2.2)-- (8.2,-2.8);
\draw (10.5,-2.5) circle (0.3);
\draw[fill = black] (9,-4) rectangle (10,-3);
\node (e) at (11,-2.5){};
\node (f) at (12,-2.5){\Huge ...};
\draw [thick, ->] (e) to (f);
\node (g) at (1.5,-5){\Huge$\vdots$};
\node (g) at (5.5,-5){\Huge$\vdots$};
\node (g) at (9.5,-5){\Huge$\vdots$};
\end{tikzpicture}
}
\caption{One segment of the safety game graph of one version of the follow game.}\label{fig:follow}
\end{figure}

In order to abstract away from the details, we turn the area in which a drone operates into a bounded two-dimensional grid world, where a number of parameters (e.g., width, height, obstacle coordinates, etc.) can be taken into account.
Every possible configuration of a specific grid world, including the positions of the robots, is modeled by a vertex in the game graph of a regular safety game. 
One snippet of such a graph for a variation of the follow game is shown in Figure~\ref{fig:follow}.
Obstacles, i.e., inaccessible grids, are marked black; the system's robot (represented by Player~$ 0 $) is depicted by a triangle. and the environment's robot (represented by Player~$ 1 $) by a circle.
A directed edge between two grid worlds indicates that there is a possible action from current configuration to reach the target configuration.
Furthermore, all parameterizations are fixed at runtime, and thus, there are no edges from a configuration into another configuration with different parameters. 

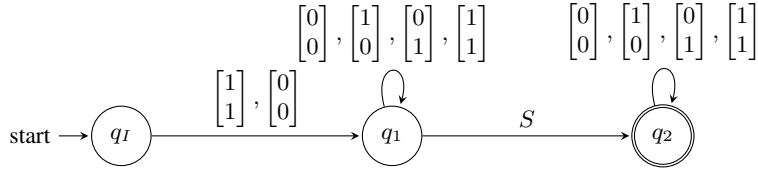
\begin{figure}
	\centering
	\begin{tikzpicture}
	\node[state, initial] at (0,0)(qi) {$q_I$};
	\node[state, right of=qi, xshift = 2.6cm](q0) {$q_1$};
	\node[state, right of=q0, xshift = 2.6cm, accepting](q1) {$q_2$};
	\draw[->]
	(qi) edge[above] node{$\begin{bmatrix}
	1 \\
	1
	\end{bmatrix},\begin{bmatrix}
	0 \\
	0
	\end{bmatrix}$} (q0)	
	(q0) edge[loop above] node{$\begin{bmatrix}
		0 \\
		0
		\end{bmatrix},\begin{bmatrix}
		1 \\
		0
		\end{bmatrix},\begin{bmatrix}
		0 \\
		1
		\end{bmatrix},\begin{bmatrix}
		1 \\
		1
		\end{bmatrix}$} (q0)
	(q0) edge[above] node{$S$} (q1)
	(q1) edge[loop above] node{$\begin{bmatrix}
		0 \\
		0
		\end{bmatrix},\begin{bmatrix}
		1 \\
		0
		\end{bmatrix},\begin{bmatrix}
		0 \\
		1
		\end{bmatrix},\begin{bmatrix}
		1 \\
		1
		\end{bmatrix}$} (q1);
	\end{tikzpicture}
    \caption{Automaton representing the grid world.\label{fig:follow-automaton}}
\end{figure}

Notice that each of the configuration in a runtime can either be ``safe'', i.e., the drone is within an acceptable proximity to the target, or ``unsafe'', i.e., beyond the proximity. 
Figure~\ref{fig:follow-automaton} shows an automaton that parameterizes the grid world of the follow game by encoding the positions of both robots as bit vectors.
The first symbol indicates which player is allowed to move their robot: 
$ \begin{bsmallmatrix}
1 \\ 1
\end{bsmallmatrix} $ 
means Player~$1$ can move their robot, whereas 
$ \begin{bsmallmatrix}
0 \\ 0
\end{bsmallmatrix} $ 
indicates Player~$0$'s turn.
The subsequent vector 
$ \begin{bsmallmatrix}
x_1 \\ x_2
\end{bsmallmatrix} $ 
encodes the $x$-coordinates of Player~$ 0 $'s and Player~$ 1 $'s robots in the unary numeral system number, respectively, followed by a separating symbol $S$ and 
$ \begin{bsmallmatrix}
y_1 \\ y_2
\end{bsmallmatrix} $ 
which encodes the $y$-coordinates. The symbol $0$ is used as padding symbol to keep the length of each word encoding a grid world to be the same.

\begin{figure}
	\includegraphics[width=1.0\linewidth,height=1\textheight]{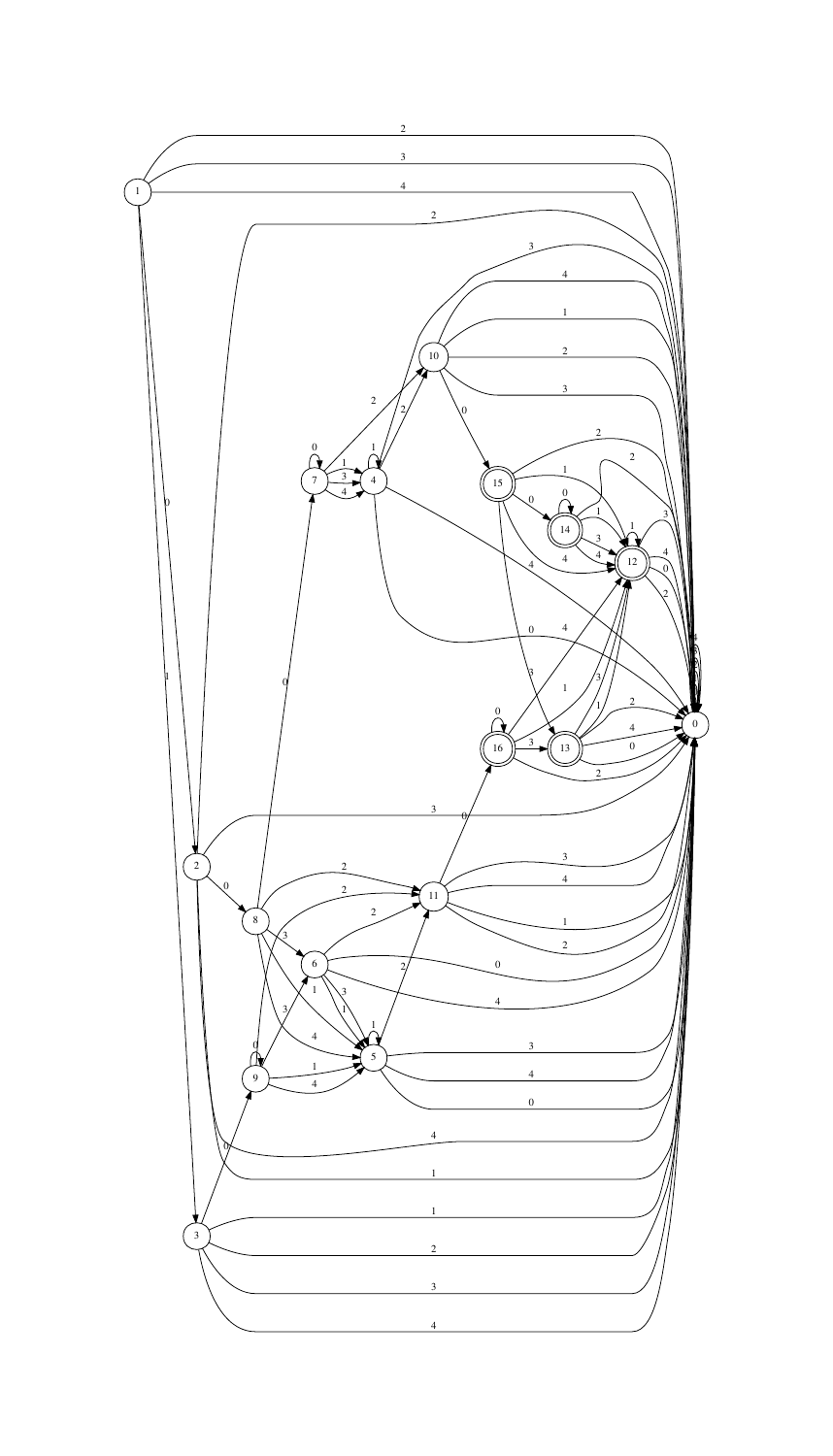}
	\caption{Automaton representing one winning strategy for a simplified
    version of the follow game. The legend for the symbols is as follows:
    $0\mapsto \protect\begin{bsmallmatrix} 1 \\ 1	\protect\end{bsmallmatrix}$, 
    $1\mapsto \protect\begin{bsmallmatrix} 0 \\ 0	\protect\end{bsmallmatrix}$, 
    $2\mapsto S$, 
    $3\mapsto \protect\begin{bsmallmatrix} 0 \\ 1	\protect\end{bsmallmatrix}$, 
    $4\mapsto \protect\begin{bsmallmatrix} 1 \\ 0	\protect\end{bsmallmatrix}$.
                    \label{fig:follow-win-automaton}}
\end{figure}

An automaton representing one winning strategy for the follow game with the robots start at the same position, and where the grid world does not contain any obstacles, is shown in Figure \ref{fig:follow-win-automaton}.
The intuition behind this automaton is that whenever Player~$1$ takes a turn, the robots are on top of each other, and once Player~$0$ takes a turn, the $x$ and $y$-coordinates differ by at most one, which translates into a simple strategy for Player~$0$: always move the robot on top of Player~$1$'s robot.
Given such a setting, the objective of the synthesis is to find a strategy that takes into account the parameters, and, regardless of the value of the parameters, works for every possible grid world.

\paragraph*{Distributed protocol example.} Consider a distributed system which operates on $n$ processes that may enter critical section. Additionally, there is a single token in the system. A process can only enter the critical section if it is in possession of the token.
We are interested in a controller which guarantees that at most one process is in the critical section at a given time. 
The controller handles the resource allocation, i.e., decides which process gets the token and how long the process keeps it. However, similar to the ring token protocol, it can only move the token to the right.
The processes can be \textit{idle} (e.g., doing computations in non-critical sections), \textit{requesting} a token, or \textit{in the critical section}.
The controller has to give a process the token if the process is in requesting state and the token passes the process.
The obvious parameter for this protocol is the amount of processes which are dependent on the system. With parameterization synthesis, it is enough to only synthesize one controller which can function regardless of the number of processes. Indeed, later in Section~\ref{sec:experiments}, we use this motivating example as one of the benchmarks---we call it ``resource allocation game''---and synthesize the controller.


\section{Preliminaries}
\label{sec:preliminaries}

Let $\mathbb N$ be the set of natural numbers.
Given two sets $A$ and $B$, we denote their \emph{symmetric difference} by $A \ominus B = (A \setminus B) \cup (B \setminus A)$.
Moreover, given a relation $E \subseteq A \times B$, the \emph{image of $A$ under $E$} is the set $E(A) = \{ b \in B \mid \exists a \in A \colon (a,b) \in E \}$; similarly, the \emph{preimage of $B$ under $E$} is the set $E^{-1}(B) = \{ a \in A \mid \exists b \in B \colon (a,b) \in E\}$.

\paragraph*{Word, Languages, and Finite Automata.}
An \emph{alphabet} is a nonempty finite set $\Sigma$ of elements, called \emph{symbols}.
A \emph{word} is a finite sequence $w = a_1 \ldots a_n$ with $a_i \in \Sigma$ for $i \in \{ 1, \ldots, n \}$.
The \emph{empty word} is the empty sequence, denoted by $\epsilon$.
The concatenation of two words $u = a_1 \ldots a_m$ and $v = b_1 \ldots b_n$ is the word $u \cdot v = a_1 \ldots a_m b_1 \ldots b_n$, abbreviated as $uv$. 
We denote the set of all words over the alphabet $\Sigma$ by $\Sigma^\ast$ and call a subset $L \subseteq \Sigma^\ast$ a \emph{language}.

A \emph{nondeterministic finite automaton (NFA)} is a tuple $\mathcal{A} = (Q, \Sigma, q_I, \delta, F)$ consisting of a nonempty finite set $Q$ of states, an input alphabet $\Sigma$, an initial state $q_I \in Q$, a transition relation $\delta \subseteq Q \times \Sigma \times Q$, and a set $F \subseteq Q$ of final states.
A \emph{run} of an NFA $\mathcal{A}$ on a word $w = a_1 \dots a_n$ is a sequence $q_0 q_1\dots q_n$ of states  such that $q_0 = q_I$ and $(q_{i-1}, a_i, q_i) \in \delta$ for $i \in \{1, \ldots, n\}$.
We call a run $q_0 \dots q_n$ accepting if $q_n \in F$.
The language of an NFA $\mathcal A$, denoted by $L(\mathcal A)$, is the set of all words $w \in \Sigma^\ast$ for which an accepting run of $\mathcal A$ on $w$ exists.
A language $L \subseteq \Sigma^\ast$ is called \emph{regular} if there exists an NFA $\mathcal A$ with $L(\mathcal A) = L$.
A \emph{deterministic finite automaton (DFA)} is an NFA where
the transition relation is effectively a function $\delta \colon Q \times \Sigma \to Q$.

A \emph{length-preserving transducer} is a tuple $\mathcal{T} = (Q, \Sigma, q_I, \delta, F)$  consisting of a nonempty finite set $Q$ of states, an input alphabet $\Sigma$, an initial state $q_I \in Q$, a transition relation $\delta \subseteq Q \times \Sigma \times \Sigma \times Q$, and a set $F \subseteq Q$ of final states.
In contrast to NFAs, which process words, a transducer processes pairs of words that have equal length (hence the name length-preserving).
More precisely, a \emph{run} of $\mathcal{T}$ on pair $(u,v) = \bigl( (a_1 \dots a_n), (b_1 \dots b_n) \bigr)$ of words is a sequence $q_0 q_1 \dots q_n$ of states such that $q_0 = q_I$ and $\bigl( q_{i-1}, (a_i, b_i), q_i \bigr) \in \delta$ for $i \in \{1,\dots, n\}$.
Similar to NFAs, the run is~\emph{accepting} if $q_n \in F$.
A transducer $\mathcal T$ defines a binary relation, denoted by $R(\mathcal T)$, that consists of all pairs
$(u, v) \in (\Sigma \times \Sigma)^\ast$ for which $\mathcal T$ has an accepting run.

\paragraph{Reactive Synthesis and Safety Games.} 
In order to synthesize controllers for reactive systems, we follow an approach popularized by McNaughton~\cite{MC93},
which translates the system and specification in question into an infinite-duration two-player game
and a controller into a winning strategy.
This approach can be easily applied to parameterized systems under suitable encoding.
Since we are interested in synthesizing systems from safety specifications,
the games we are faced with are so-called \emph{safety games}~\cite{gradel2002automata}.
The basic building block of a safety game is an \emph{arena} $\mathcal{A} = (V_0, V_1, E)$,
which is a directed graph with a countable vertex set $V = V_0 \uplus V_1$ and directed edge relation $E \subseteq V \times V$.
The game has two players:
\emph{Player~0}, who represents the system, controls the vertices in $V_0$,
and \emph{Player~1}, who represents the environment, controls the vertices in $V_1$.

Formally, a \emph{safety game} is a triple $\mathcal{G} = (\mathcal{A}, I, B)$ consisting of an arena $\mathcal A = (V_0, V_1, E)$, a set $I \subseteq V$ of initial vertices, and a set $B \subseteq V$ of bad vertices.
A safety game is played as follows: initially, a token is placed on one initial vertex $v_0 \in I$; then, the player having control over the vertex moves the token along one of the outgoing edges to the next vertex. 
The process of moving the token is repeated ad~infinitum, resulting in an infinite sequence $\pi = v_0v_1\ldots$ of vertices where $v_0 \in I$ and $(v_i, v_{i+1}) \in E$ for all $i \in \mathbb{N}$.
We call such a sequence a \emph{play}.

In a safety game, Player~0's goal is to keep the token away from the bad vertices, while Player~1's goal is to reach them.
Formally, a play $\pi = v_0v_1\dots$ is \emph{winning for Player~0} if $v_i \notin B$ for all $i \in \mathbb{N}$.
Conversely, it is winning for Player~1 if $v_i \in B$ for some $i \in \mathbb{N}$.
Hence either Player~1 or Player~2 wins for each play.

In McNaughton's framework, synthesizing a controller amounts to computing a so-called winning strategy for Player~$0$.
Formally, a \emph{strategy} for Player~$0$ is a mapping $\sigma \colon V^\ast \times V_0 \to V$ such that $\bigl(\sigma(v_0 \dots v_n), v_n \bigr) \in E$ for every finite play prefix $v_0 \dots v_n \in V^\ast V_0$.
We say that a play $\pi = v_0v_1\dots$ is \emph{played according to $\sigma$} if $v_i = \sigma(v_0 \ldots v_{i-1})$ for every $i \in \mathbb N$ such that $v_i \in V_0$.
Moreover, a strategy is said to be \emph{winning} if every play that is played according to $\sigma$ is winning.

In this paper, we do not compute winning strategies directly but instead learn a proxy object, called \emph{winning set}.
Intuitively, a winning set is a set $W \subseteq V$ of vertices that contains all initial vertices, contains no bad vertex, and is a ``trap'' for Player~1 in the sense that Player~1 cannot force the play to a vertex outside the winning set.
Formally, winning sets are defined as follows.

\begin{definition}[Winning set]\label{defn:winning-set}
Let $\mathcal{G}=(\mathcal{A},I,B)$ be a safety game over the arena $\mathcal{A} = (V_0, V_1, E)$.
A \emph{winning set} is a set $W \subseteq V$ of vertices satisfying the following four properties:
\begin{enumerate}
	\item $I \subseteq W$: all initial vertices are subsumed by the winning set (\emph{initial condition}).
	\item $B \cap W = \emptyset$: no bad vertex is contained in the winning set (\emph{bad condition}).
	\item $E(\{v\}) \cap W \neq \emptyset$ for all $v \in W \cap V_0$: every vertex of Player~0 inside the winning set has at least one outgoing edge connected to another vertex inside the winning set (\emph{existential closedness}).
	\item $E(\{v\}) \subseteq W$ for all $v \in W \cap V_1$: the successors of every Player~1 vertex inside the winning set is also inside the winning set (\emph{universal closedness}).
\end{enumerate}
\end{definition}

A winning strategy for Player~0 can be derived from a winning set $W$ in a straightforward manner:
starting with a vertex $v \in I$ (and, hence, $v \in W$), every time Player~0 is in control of the token, the strategy is to move the token to a successor vertex which is also inside the winning set $W$.
It is not hard to verify that this strategy is in fact winning for Player~0 from every vertex in $W$:
first, all initial vertices are contained in the winning set, and every Player~0 vertex has a successor which is inside the winning set; 
second, since Player~1 can never leave the winning set (due to universal closedness) and since no vertex inside the winning set is bad, it is guaranteed that following the strategy results in a winning play regardless of the moves of Player~1.
\paragraph*{Regular safety games.}
We represent safety games using finite automata and transducers.
A \emph{regular arena} is an arena $\mathcal{A}_{\mathcal{R}} = (L(\mathcal{A}_{V_0}), L(\mathcal{A}_{V_1}), R(\mathcal{T}_E))$ where $\mathcal{A}_{V_0}$ and $\mathcal{A}_{V_1}$ are NFAs and $\mathcal{T}_E$ is a length-preserving transducer.
A \emph{regular safety game} is a safety game $\mathcal{G}_{\mathcal{R}} = (\mathcal{A}_{\mathcal{R}}, L(\mathcal{A}_I), L(\mathcal{A}_B))$ where $\mathcal{A}_I$ and $\mathcal{A}_B$ are given as NFAs.


\section{Algorithm}
\label{sec:algo}

\begin{figure}[t]
	\centering
	\scalebox{0.85}{
		\begin{tikzpicture}[punkt/.style={rectangle, rounded corners,
			draw=black, very thick, text width=5em,
			minimum height=8em, text centered}]
		\node[punkt] (learner) {};
		\node[align=left] (learner_text) at ($(learner) + (0, 1cm)$) {Learner};
		\node[punkt, minimum height=12em, text width=11em, right=3.5cm of learner] (teacher) {};
		\node[align=left] (teacher_text) at ($(teacher) + (0, 1.6cm)$) {Teacher};
		\node[rectangle, minimum height=2em, rounded corners, draw=black, text centered] (teacher_mem) at ($(teacher)+(0,2.5em)$) {$w\in R?$};
		\node[align=left, rectangle, minimum height=6em, rounded corners, draw=black] (teacher_equ) at ($(teacher)+(0,-2.5em)$) {$R=L(\mathcal{A}_H)?$};
		
		\draw ($(learner.east)+(0, 3em)$)
		edge[->] node[above,pos=0.4] { $Mem(w)$ }
		($(teacher_mem.west)+(0, 0.5em)$);
		\draw ($(learner.east)+(0, 2em)$)
		edge[<-] node[below,pos=0.4] { $\mathit{yes}$ or $\mathit{no}$ }
		($(teacher_mem.west)+(0, -0.5em)$);
		\draw ($(learner.east)+(0,-2em)$)
		edge[->] node[above,pos=0.43] { $Eq(\mathcal{A}_H)$ }
		($(teacher_equ.west)+(0,+0.5em)$);
		\draw ($(learner.east)+(0,-3em)$)
		edge[<-] node[below,pos=0.43] { $\mathit{yes}$ or $(\mathit{no}, w)$ }
		($(teacher_equ.west)+(0,-0.5em)$); ;
		
		\end{tikzpicture}
	}
	\caption{General active automata learning framework. The teacher must be able to answer $w \in R?$ and must have some way to determine whether $R=L(\mathcal{A}_H)$.}
	\label{fig:active-learning}\vspace{0.1cm}
\end{figure}
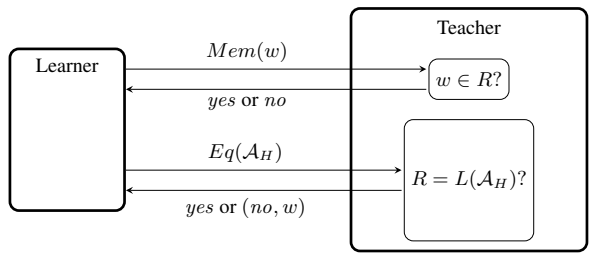

\subsubsection{An active automata learning algorithm}

Suppose $R$ is a regular language whose definition is not directly accessible.
\emph{Automata learning}
algorithms~\cite{angluin1987learning,rivest:inference1993,learning-book,bollig:angluin2009}
automatically infer a DFA $\mathcal{A}_H$ recognising~$R$.
The setting of an active learning algorithm is shown in Figure~\ref{fig:active-learning}
assumes a~\emph{teacher} who has access to $R$ and can answer the
following two queries: (1) Membership query $Mem(w)$: is the word $w$ a~member of $R$, i.e., $w \in R$?
(2) Equivalence query $Eq(\mathcal{A}_H)$: is the language of~$\mathcal{A}_H$ equal to
$R$, i.e., $L(\mathcal{A}_H) = R$?
If not, it returns a counterexample $w \in L(\mathcal{A}_H) \ominus R$.
The learning algorithm will then construct an DFA $\mathcal{A}_H$ such
that $L(\mathcal{A}_H) = R$ by interacting with the teacher.
Such an algorithm works iteratively: in each iteration, it
performs membership queries to get from the teacher information about $R$.
Using the results of the queries, it proceeds by
constructing a~hypothesis DFA $\mathcal{A}_H$ and makes an
equivalence query $Eq(\mathcal{A}_H)$. If $L(\mathcal{A}_H) = R$, the learning algorithm terminates and outputs $\mathcal{A}_H$.
Otherwise, the algorithm uses the counterexample $w$
returned by the teacher to refine the hypothesis DFA in the next iteration.

For completeness, we briefly describe how the learning algorithm computes hypothesis automata.
The foundation of the algorithm is the Myhill-Nerode theorem~\cite{nerode1958linear},
which states that the minimal DFA recognizing $R$
is isomorphic to the set of equivalence classes defined by the following relation:
$x \equiv_R y\ \mbox{iff it holds that}\ \forall z\in \Sigma^*: xz\in R \leftrightarrow yz\in R.$
Informally, two words $x$ and $y$ belong to the same state of the minimal DFA recognising $R$ iff they cannot be distinguished by any suffix $z$. In other words, if one can find a suffix $z'$ such that $xz' \in R$ and $yz' \notin R$ or vice versa, then $x$ and $y$ belong to different states of the minimal DFA.

The learning algorithm maintains a Boolean table
where the rows are indexed by $X\subseteq \Sigma^*$ and the columns indexed by $Y\subseteq \Sigma^*$.
Each cell $(x,y)$ of the table indicates whether or not $xy \in R$.
For $x,x'\in X$, we write $x \sim_Y x'$  iff $xy \equiv_R x'y$ for all $y \in Y$.
Note that $\sim_Y$ is an equivalence relation over $X$, and
that $x \sim_Y x'$ iff the rows indexed by $x$ and $x'$ contain the identical Boolean values.
The table is \emph{consistent} iff for all $x,x'\in X$ and $x \neq x'$, it holds that $x \not\sim_Y x'$.
The table is \emph{closed} iff for all $x\in X$ and $a \in \Sigma$, there exists $x' \in X$ such that $xa \sim_Y x'$.
By the Myhill-Nerode theorem, the table determines a DFA when it is consistent and closed:
the states of the DFA are $\{[x]_Y : x \in X\}$
(where $[\cdot]_Y$ is the equivalence classes induced by $\sim_Y$),
the accepting states are $\{[x]_Y : x \in X \cap R \}$,
and the transition function $\delta : [X]_Y \times \Sigma \to [X]_Y$ is defined by $\delta ([x]_Y, a) = [xa]_Y$.
Note that this DFA is minimal as every two states of it can be distinguished by some word in $Y$
by the definition of consistency.

During the learning process, the algorithm fills and extends the table through membership queries
until the table is consistent and closed.
The algorithm then determines a hypothesis automaton $\mathcal{A}_H$ from the table
and makes an equivalence query $Eq(\mathcal{A}_H)$.
If the teacher returns a counterexample $w$, 
the algorithm will perform a binary search over $w$ using membership queries to find a suffix $y$ of $w$
and extend $Y$ to $Y\cup \{y\}$, which will identify at least one more state for $R$ by the Myhill-Nerode theorem.

\begin{proposition}[\cite{rivest:inference1993}]\label{thm:lstar}
    The learning algorithm in Figure~\ref{fig:active-learning} finds
    the minimal DFA $\mathcal{A}_H$ for the target regular language $R$ using at most $n$ equivalence queries and $n(n+n|\Sigma|) + n\log m$ membership queries, where $n$ is the number of state of $H$ and $m$ is the length of the longest counterexample returned from the teacher.
\end{proposition}


\subsubsection{A teacher for learning winning set}

Let $\mathcal{G}_{\mathcal{R}} = (\mathcal{A}_{\mathcal{R}}, L(\mathcal{A}_I), L(\mathcal{A}_B))$
be a regular safety game with regular arena 
$\mathcal{A}_{\mathcal{R}} = (L(\mathcal{A}_{V_0}), L(\mathcal{A}_{V_1}), R(\mathcal{T}_E))$.
We describe below a teacher to learn a regular winning set for $\mathcal{G}_{\mathcal{R}}$.
Since $\mathcal{G}_{\mathcal{R}}$ can have multiple winning sets,
we aim to learn the \emph{maximal} winning set,
which, if exists, is unique as winning sets are closed under union.
\begin{theorem}
	The target object in Figure \ref{fig:active-learning}, the maximal winning set, is unique.
\end{theorem}
\paragraph*{Membership query.}
To answer a membership query $Mem(w)$,
the teacher needs to check whether Player~$1$ can force Player~$0$ to visit a bad vertex from vertex $w$.
Since the transition relation is length-preserving, only a finite number of vertices
(i.e.~at most $|\Sigma|^{|w|}$ vertices) can be reached from vertex $w$.
Therefore, this check can be done by solving an induced \emph{finite} safety game with
$I_w = \{w\}$ as the set of initial vertices
and $B_w = \{w' \in L(\mathcal{A}_B) : |w'| = |w| \}$ as the set of bad vertices.
Safety games over finite graphs are known to be decidable~\cite{gradel2002automata},
thus making our membership query decidable.
\paragraph*{Equivalence query.}
To answer an equivalence query $ Eq(\mathcal{A}_H)$, the teacher simply checks that all conditions in
Definition~\ref{defn:winning-set} are fulfilled by the hypothesis DFA $\mathcal{A}_H$.
Note that a DFA satisfying these conditions serves as a proof for safety
even if it does not recognize the maximal winning set.
The pseudo code of the equivalence check can be found in Algorithm \ref{alg:equiv-check}.
Given an equivalence query $Eq(\mathcal{A}_H)$ by the learner, the teacher first checks if $L(\mathcal{A}_I) \not \subseteq L(\mathcal{A}_H)$ and if there is $v \in L(\mathcal{A}_I) \setminus L(\mathcal{A}_H)$, the teacher returns $v$ as a counterexample.

Secondly, the teacher checks whether $ L(\mathcal{A}_B) \cap L(\mathcal{A}_H) \neq \emptyset$. If there is a $v \in L(\mathcal{A}_B) \cap L(\mathcal{A}_H)$, then the teacher returns $v$ as a counterexample.

According to the third part of Definition \ref{defn:winning-set}, the teacher checks if there exists $v \in L(\mathcal{A}_H) \cap L(\mathcal{A}_{V_0})$ and $R(\mathcal{T}_E)(\{v\}) \cap L(A_H) = \emptyset$. Here either $v$ should be excluded from the hypothesis or one of its successors should be included. The teacher then makes membership queries to check if $v$ should be excluded: if $Mem(v)$ returns ``no'', the teacher returns $v$ as counterexample. Otherwise, the teachers returns some $u \in R(\mathcal{T}_E)(\{v\})$ as a counterexample such that $Mem(u)$ is ``yes''.

Lastly, the teacher checks if there exists $v \in L(\mathcal{A}_H) \cap L(\mathcal{A}_{V_1})$ and $R(\mathcal{T}_E)(\{v\}) \not \subseteq L(\mathcal{A}_H)$. Again, either $v$ should be excluded or one of its successors should be included. If $Mem(v)$ returns ``no'', the teacher returns $v$ as a counterexample. Otherwise, the teacher returns some $u \in R(\mathcal{T}_E)(\{v\}) \setminus L(\mathcal{A}_H)$ as a counterexample.

Since the teacher checks all conditions in Definition~\ref{defn:winning-set} for an equivalence query,
if the teacher replies ``yes''  then the hypothesis DFA indeed recognizes a winning set.
Otherwise, the teacher will pinpoint a counterexample violating the definition.
Furthermore, observe that the counterexamples pinpointed by the teacher
are located in the symmetric difference of the candidate language and the maximal winning set.
Therefore, if the maximal winning set can be recognized by a DFA of $n$ states,
the learning algorithm will terminate in $n$ iterations by Proposition~\ref{thm:lstar}.
We summarize the soundness and completeness of our learning method in the following theorem.

\begin{theorem}
	Given a regular safety game $\mathcal{G}_{\mathcal{R}} = (\mathcal{A}_{\mathcal{R}}, L(\mathcal{A}_I), L(\mathcal{A}_B))$, 
	the learning algorithm in Figure~\ref{fig:active-learning} computes a winning set on termination.
	Furthermore, when the maximal winning set $W$ is regular, the algorithm will terminate in at most $n$ iterations
	where $n$ is the size of the minimal DFA of $W$.
\end{theorem}

%
%
%

\begin{algorithm}[t]
	\DontPrintSemicolon
	\KwIn{$\mathcal{G}_{\mathcal{R}} = (\mathcal{A}_{\mathcal{R}}, L(A_I), L(A_B))$  over the regular arena $\mathcal{A}_{\mathcal{R}} = (L(A_{V_0}), L(A_{V_1}), R(\mathcal{T}_E))$ and an hypothesis DFA $\mathcal{A}_H$. }
	\BlankLine
	\uIf{$L(A_I) \setminus L(A_H) \neq \emptyset$}{ Find some $v \in L(A_I) \setminus L(A_H)$ and \textbf{return} $(\mbox{``no''},v)$ }
	\BlankLine
	\uIf{$L(A_H) \cap L(A_B) \neq \emptyset$}{ Find some $v \in L(A_H) \cap L(A_B)$ and \textbf{return} $(\mbox{``no''},v)$}
	\BlankLine
	\uIf{there is $v \in L(A_{V_0}) \cap L(A_H)$ such that $R(\mathcal{T}_E)(\{v\}) \cap L(A_H) = \emptyset$}
	{\uIf{$Mem(v)$ is ``yes''}{
			{Find some $u \in R(\mathcal{T}_E)(\{v\})$ such that $Mem(u)$ is ``yes''\; \textbf{return} $(\mbox{``no''},u)$}
		}\uElse{
			\textbf{return} $(\mbox{``no''},v)$
	}}
	\BlankLine
	\uIf{there is $v$ such that $v \in L(A_{V_1}) \cap L(A_H)$ and $R(\mathcal{T}_E)(\{v\}) \not \subseteq L(A_H)$}
	{\uIf{$Mem(v)$ is ``yes''}{
			Find some $u \in R(\mathcal{T}_E)(\{v\}) \setminus L(A_H)$ and \textbf{return} $(\mbox{``no''},u)$
		}\uElse{
			\textbf{return} $(\mbox{``no''},v)$
	}}
	\BlankLine
	{\textbf{return} ``yes''}
	\caption{Resolving an equivalence query for regular safety games} \label{alg:equiv-check}
\end{algorithm}


\section{Case Studies and Experiments}
\label{sec:experiments}
In this section, we provide some case studies as benchmarks and report the results of the experiments based on given benchmarks. In order to asses the performance of our tool, \textit{\PLSynth}, we compare it with three existing tools that are able to solve safety games over infinite graphs: \textit{SAT-Synth}, \textit{RPNI-Synth}~\cite{NT16} and \textit{DT-Synth}~\cite{NM19}

\paragraph*{Tools.}
The tools \textit{SAT-Synth}\xspace and \textit{RPNI-Synth}\xspace both compute a winning set based on learning finite automata with a teacher that answers to equivalence queries.
In contrast to \textit{\PLSynth}---which solves regular safety games---these tools are able to solve \textit{rational safety games}, which is a more general type of safety games, since in these games, edge relations may be represented by non length-preserving transducers.
Furthermore, the learner of \textit{SAT-Synth}\xspace uses a SAT solver to learn automata, while  \textit{RPNI-Synth}\xspace is based on the popular RPNI learning algorithm \cite{OG92}.

The tool \textit{DT-Synth} uses formulas in the first-order theory of linear integer arithmetic to encode safety games. It uses a learning algorithm that learns from data in the form of Horn clauses. The teacher in this tool was built on top of the constraint solver Z3 \cite{Z3}.

\textit{\PLSynth}\xspace is implemented with the use of automata libraries and an existing implementation of an \Lstar\ learner\cite{CHLR17}.
The teacher is implemented in Java and uses existing automata methods to
implement the algorithms from Section \ref{sec:algo}.
The input format is a text file which encodes a regular safety game $\mathcal{G}_{\mathcal{R}} = (\mathcal{A}_{\mathcal{R}}, L(\mathcal{A}_I), L(\mathcal{A}_B))$.\footnote{Code and benchmarks are available at https://github.com/lstarsynth/lstar-psynth.}

 The teacher for \textit{\PLSynth} is an extension of the one used by \textit{SAT-Synth}, \textit{RPNI-Synth}, and \textit{DT-Synth}: it also answers to membership queries in order to accommodate for the additional queries the learner might ask, since, beside equivalence queries, our learner also asks membership queries.

\paragraph*{Benchmarks.}
Some of the benchmarks are taken from \cite{NT16} with some modification to fit the framework of regular safety games. In particular, we adjust the arenas of the game, from infinite arenas into arenas with arbitrary but bounded size.
The other benchmarks are either known games which are translated to a regular safety game, e.g., the Nim game \cite{Bou1901}, or inspired by some processes that happen in real world, such as resource allocation protocols or the movement of an autonomous robotic vacuum cleaner.
The list of benchmarks is as follows:
\begin{description}
	\item [Box game:] A robot moves in an two-dimensional grid world of size $n \times m $ with $n,m \geq 3$.\footnote{The encoding in the benchmarks use a grid world of size $2^n \times 2^n$ which can be easily reduced to $n \times m $} Player~$0$ controls the vertical movement of the robot while Player~$1$ controls the horizontal movement. Player~$0$ wins if the robot stays within a horizontal stripe of width 3 around the middle of the arena. We can consider this kind of game as an abstraction of some autonomous control system, e.g., a controller that ensures a drone stay in some range of altitude.
	\item [Control unit game:] Consider a system that controls the temperature of $ n $ power plants within a certain safe level. We can model this as a game between two players, $0$ and $1$. Player~$0$ acts as the controller who can decrease the temperature of some plant (e.g., by reducing the boiler temperature.) Player~$1$ acts as the environment who may increase the temperature of some plant (e.g., weather changes, cooling system malfunction). The game is played in a sequential fashion, i.e., Player~$0$ and Player~$1$ can alternately increase or decrease the temperature of a plant. Player~$0$ wins if none of the plants reach critical temperature.
	\item[Diagonal game:] A variation of the Box game where Player~$0$ again controls the vertical movement and Player~$1$ controls the horizontal movement of a robot in a bounded two-dimensional grid world. Player~$0$ wins if the robot stays within a two cells of the diagonal in the arena.
	\item[Evasion game:] Two robots are moving in an bounded discrete two-dimensional grid world of size  $n \times m$ with $n,m \geq 3$. Each Player is in control of one robot and they can move their respective robot at most one cell in any direction (either vertically or diagonally.) If the system moves its robot outside of a bound it automatically wins\footnote{The original version of the evasion game is played in an infinite grid world, thus, making one valid strategy to always move into one direction, which resembles Player~$0$ moving out of bound.}. Player~$0$  wins if Player~$1$ never moves its robot on top of Player~$0$'s robot.	
	\item[Follow game:] A variation of the evasion game where Player~$0$ wins if it manages to keep its robot within a Manhattan distance of two cells to Player~$1$'s robot.
	\item [Nim game:] The standard Nim game consists of three piles of chips and two players taking alternating turns. On each turn, each player must remove one chip, and may remove any number of chips so long as they all come from the same pile. The player who removes the last chip wins the game\footnote{This version of winning condition is called ``mis\`ere play condition'', in which the last player making a move loses. Nim can also be played with ``normal play condition'', i.e., the last player making a move wins.}. The game is modified to be an infinite duration game by adding an infinite loop at the end of the game. A winning strategy is computed for all winning starting positions which are determined by the \textit{Nim sum}. More information on the Nim game and its winning strategy can be found in \cite{TSF2014}.
	\item [Resource allocation game:] This game involves a single token and $n$ processes. Each process has three states: \textit{idle}, \textit{requesting}, and \textit{in critical section}. A process can move from a requesting state to the critical section if and only if it has the token. If a process is in a requesting state, it is guaranteed by design of the game, that it will eventually get the token. Player~$0$ controls the token and can either: (i) move the token from one process to another, or (ii) keep it in the same place if the process is in the critical section, or if there are only idle processes. Player 1 can change the state of a process from idle to requesting or vice versa. Additionally, Player 1 can move a process to the critical section if the process is in control of the token. Once a process enters the critical section, it may stay in the critical section even without the token. Player 0 wins if at all times, there is no process in the critical section without the token.
	\item [Robot vacuum cleaner game:]  A vacuum cleaner robot and a human move in an two-dimensional grid world of size $2^n \times 2^n$ with $n \geq 2$. Player 0 controls the movement of the robot and Player~$1$ controls the movement of the human. Player~$0$ wins if the robot never bumps into the human, and if the human tries to step on the robot, it moves away.
	\item [Solitary box:] Another variation of the Box game where only Player~$0$ controls the vertical and horizontal movement of the robot.
\end{description}
\begin{table*}[t]
	\caption{
		Results on the benchmarks on \textit{\PLSynth}\xspace, \textit{SAT-Synth}\xspace and \textit{RPNI-Synth}\xspace. ``Size'' measures the size of the final automata synthesized by the algorithms. ``---'' indicates a timeout after 300s. ``N/A'' corresponds to not supported by the tool.
	} \label{tab:benchmarks}
	
	\centering
	\resizebox{\columnwidth}{!}{%
		\begin{tabular}{lrr@{\hskip 2em}rr@{\hskip 2em}rr@{\hskip 2em}r}
			\toprule
			& \multicolumn{2}{c}{\hskip -1em \textit{\PLSynth}\xspace} & \multicolumn{2}{c}{\hskip -1em \textit{SAT-Synth}\xspace} & \multicolumn{2}{c}{\hskip -1em \textit{RPNI-Synth}\xspace} &\textit{DT-Synth}\xspace \\ 
			
			\cmidrule(lr{2em}){2-3} \cmidrule(l{-0.25em}r{2em}){4-5} \cmidrule(l{-0.25em}r{2em}){6-7}  \cmidrule(lr){8-8}
			Game  & Time in s & Size &Time in s& Size & Time in s&  Size & Time in s\\
			\midrule
			Box &$ 1.62$ & $5$ & $6.83$ &$4$ & $1.92$ & $7$ & $5.76$ \\
			Control unit &$ 0.40$ & $3$& $185.50$ & $5$ & $1.13$ &$5$ & N/A  \\
			Diagonal &$ 0.68$ & $3$ & $113.52$ &$7$ & $1.62$ & $7$ & $139.36$ \\
			Evasion &$ 4.77$ & $11$  & $122.41$ & $7$ & $2.52$ &$11$ & $10.83$ \\
			Follow &$ 6.71$ & $16$ & $207.12$ & $16$& $18.53$ &$16$  & $31.67$ \\
			Nim &$ 3.64$ & $4$ & \multicolumn{2}{c}{------------} & $7.12$ & $5$  & N/A\\
			Resource allocation &$ 0.65$ & $4$ & $24.00$ & $3$ & $3.77$ & $4$ & N/A\\
			Robot vacuum cleaner &$ 1.21$ & $3$ &\multicolumn{2}{c}{------------} & \multicolumn{2}{c}{------------} & --------- \\
			Solitary box &$ 1.14$ & $4$ & $5.71$ &$4$ & $0.30$ & $4$  & $1.89$ \\
			\bottomrule
		\end{tabular}%
	}
\end{table*}

\paragraph*{Results.}
The result of the benchmarks on \textit{\PLSynth}\xspace, \textit{SAT-Synth}\xspace, \textit{RPNI-Synth}\xspace and \textit{DT-Synth} is shown in Table \ref{tab:benchmarks}. 
In this table, we report the time each tool took to synthesize an automaton that encodes a winning set, as well as the size of the respective automaton\footnote{Apart from \textit{DT-Synth}, since instead of automata, it produces witnesses as decision trees.}.
We conducted the experiments on an Intel Xeon E7-8857 v2 CPU with 4 GB of RAM running a 64-bit Debian operating system. From the results, we can see that 
\textit{\PLSynth}\xspace was able to solve all games, whereas \textit{RPNI-Synth} and \textit{DT-Synth} were not able to solve the robot vacuum cleaner game, and \textit{SAT-Synth}\xspace did not solve the robot vacuum cleaner game and the Nim game.
Moreover, the aggregated runtime to solve all 9 games for \textit{\PLSynth}\xspace is 20.82 seconds compared to \textit{RPNI-Synth}\xspace which took 36.91 seconds to solve 8 games in total. \textit{SAT-Synth}\xspace was able to solve 7 games taking 665.09 seconds.
Finally, \textit{DT-Synth} was only able to solve 5 games within 189.51 seconds---this is partly due to the inability of \textit{DT-Synth} encoding to represent three benchmarks: control unit, Nim, and resource allocation.
Given the results, it is not surprising that \textit{\PLSynth}\xspace was able to outperform the other tools, since the benchmarks are more well suited for regular safety game framework.
On the other hand, if we consider the size of the solutions, \textit{RPNI-Synth} performed worst, with only 2 out of 9 solutions that are at least as small as those produced by other tools, followed by \textit{SAT-Synth}\xspace 5 out of 9 games. \textit{\PLSynth} performed best with 6 out of 9 solutions that are at least as small as others\footnote{Including one case (robot vacuum cleaner) in which the other two tools timed out.}.
Again, this is not a surprising result with respect to \textit{RPNI-Synth} performance, since it was not tailored to find small solutions, whereas \textit{SAT-Synth} was designed to find such solutions. However, although \textit{\PLSynth} was also not tailored to optimize the solution size\footnote{In spite of the fact that Angluin's algorithm computes the minimal DFA for a given target language, it is not necessarily encoded by a small automaton.}, it produced better solutions compared to \textit{SAT-Synth}.
From the experiments, it appears that \textit{\PLSynth}\xspace performs well on benchmarks where a winning strategy can be synthesized by only looking at small $n$ in the parameterization. 
If larger $n$ is needed in order to find a winning strategy, the runtime significantly increases (up to 5-10 times as much time needed) as in the case for the evasion, follow and Nim game.
We believe this correlates to the runtime of Angluin's algorithm which is strongly dependent on the length of words and counterexamples considered in a given run, which increases as $n$ increases.

\paragraph*{Parameterization in DT-Synth.}
Encoding the benchmarks as safety games in \textit{DT-Synth} is not straightforward, and, in some cases, not possible (i.e., with control unit, Nim, and resource allocation.)
This is because, in those corresponding cases, either the games specifically parameterize the amount of processes, or perform bit-sensitive operations. For the rest of the games that are played on arenas of the size $n \times m$, this can be represented in \textit{DT-Synth} by letting the environment pick two additional variables, $n$  and $m$.
These variables further constrain the initial states and modify the transition system accordingly, i.e., enable/disable transitions, based on their value.

\section{Related work}
In the context of safety games, 
a constraint-based approach for solving safety games over infinite graphs \cite{BCPR14,Katis18} and various learning approaches for finite graphs and infinite graphs have been proposed \cite{NM19,NT16,N11}.
Similar to the framework of Neider et al. \cite{NT16} we encode safety games symbolically using the idea of regular model checking. Their work considers rational safety games which differ with our regular safety games in the definition of the edge relation. 
The edge relation in our framework is encoded by length-preserving transducers while rational safety games allow a more general type of transducer.
The framework for solving rational safety games is implemented in two tools, \textit{SAT-Synth} and \textit{RPNI-Synth}.
On the other hand, the framework in another learning-based approach, which is implemented in the tool \textit{DT-Synth}, does not fix the representation of safety games and uses formulas in the first-order theory of linear integer arithmetic to encode them \cite{NM19}.
This leads to some encoding difficulties with parameterized systems as discussed in Section \ref{sec:experiments}.
The learner in both frameworks learns passively from a sample and can only ask the teacher equivalence queries while the algorithm we design is able to employ a learner which is allowed to ask membership queries in addition to equivalence queries.
All frameworks mentioned above operate on safety games over infinite-state arenas, whereas we consider infinitely many finite graphs due to the nature of length-preserving transducers.
However, this is not a restriction as we can parameterize the value that goes towards infinity and finding a strategy which works for every $n$ also gives us a strategy for every specific place in the infinite-state arena for an appropriately chosen $n$.
There might be games which will not have a strategy for finite graphs (see evasion game in Section \ref{sec:experiments}) where we extend transitions to go ``out of bound'' of the parameter and always stay safe.
This works because there is a way for one robot to catch the other then there is going to be a finite example on grid world with a specific size.

The framework of regular model checking is used in many different areas of research to verify different properties such as safety \cite{CHLR17,DBLP:journals/entcs/HabermehlV05,NT16,NJ13} or liveness \cite{LR16,PS00,FITPUB8352}. In particular, for verification of those properties in parameterized systems regular model checking has seen successful application \cite{CHLR17,LR16}.
Furthermore, the approaches in \cite{CHLR17,LR16} also employ Angluin-style \Lstar-learning to verify properties of parameterized systems.

\section{Conclusion}

\label{sec:conclusion}

In this paper, we have developed a learning-based methodology for synthesizing parameterized systems from safety specifications.
Our approach reduces this synthesis problem to a two-player safety game in an infinite arena, where synthesizing a controller amounts to computing a winning strategy (a winning set) for the player embodying the system.
Inspired by Regular Model Checking and the work by Neider and Topcu, we encode sets of vertices by means of finite automata and edges using length-preserving transducers.
This encoding allows us to utilize Angluin's popular automata learning algorithm, which significantly reduces the complexity of the underlying learning problem as compared to the earlier work by Neider and Topcu (the former being polynomial while the latter being NP-complete).
In fact, our experimental evaluation shows that a prototype of our approach is very effective in synthesizing various types of parameterized systems, including process resource allocation and robotic motion planning.


There exist various interesting directions for future work.
First, we plan to extend our framework to liveness properties, for example, by learning \textit{ranking functions} rather than winning sets \cite{FPPZ2004a,FPPZ2004b}.
Second, we would like to consider game arenas with uncountably many vertices, which often arise in the context of cyber-physical systems.
One possible approach to this problem would be to encode such arenas by means of $\omega$-regular languages and $\omega$-transducers, and then use existing learning algorithms for $\omega$-automata (e.g., Büchi automata) to learn winning sets~\cite{DBLP:journals/tcs/AngluinF16}.
Finally, we want to modify our approach such that it learn a strategy directly rather than a proxy object (i.e., a winning set).
This would allow us to also optimize for other criteria such as size or number of operations required to compute the next move.

\section*{Acknowledgement}
This work was partially funded by the ERC Starting Grant
AV-SMP (grant agreement no. 759969) and MPI-Fellowship as well as the DFG grant no. 434592664.
%
%
\bibliographystyle{splncs04}
\bibliography{bib}

\end{document}